\documentclass[epj]{webofc}
\usepackage[utf8]{inputenc}
\usepackage[varg]{txfonts}   
\usepackage{booktabs}
\usepackage{xcolor}
\definecolor{darkred}{rgb}{0.4,0.0,0.0}
\definecolor{darkgreen}{rgb}{0.0,0.4,0.0}
\definecolor{darkblue}{rgb}{0.0,0.0,0.4}
\usepackage[bookmarks,linktocpage,colorlinks,
    linkcolor = darkred,
    urlcolor  = darkblue,
    citecolor = darkgreen]{hyperref}
\usepackage{subfigure}
\usepackage{braket}
\usepackage{latexsym}
\usepackage{epstopdf}
\usepackage{grffile}

\wocname{EPJ Web of Conferences}
\woctitle{Lattice2017}
\newcommand{\lp}{\left(}
\newcommand{\rp}{\right)}

\newcommand{\del}{\partial}

\newcommand{\calJ}{\mathcal{J}}
\newcommand{\calK}{\mathcal{K}}

\newcommand{\calO}{\mathcal{O}}
\newcommand{\real}{\mathbb{R}}
\newcommand{\complex}{\mathbb{C}}

\renewcommand{\Re}{\mathrm{Re}}
\renewcommand{\Im}{\mathrm{Im}}
\begin{document}
%
\selectlanguage{english}
\title{%
On a modification method of Lefschetz thimbles 
}
\author{%
\firstname{Shoichiro} \lastname{Tsutsui}\inst{1}\fnsep\thanks{Speaker, \email{stsutsui@post.kek.jp}} \and
\firstname{Takahiro M.} \lastname{Doi}\inst{2}
}
\institute{%
KEK Theory Center, High Energy Accelerator Research Organization,
1-1 Oho, Tsukuba, Ibaraki 305-0801, Japan
\and
Theoretical Research Division, Nishina Center, RIKEN, Wako 351-0198, Japan
}
\abstract{%
  The QCD at finite density is not well understood yet, where standard Monte Carlo simulation suffers from the sign problem. In order to overcome the sign problem, the method of Lefschetz thimble has been explored. Basically, the original sign problem can be less severe in a complexified theory due to the constancy of the imaginary part of an action on each thimble. However, global phase factors assigned on each thimble still remain. Their interference is not negligible in a situation where a large number of thimbles contribute to the partition function, and this could also lead to a sign problem.
  In this study, we propose a method to resolve this problem by modifying the structure of Lefschetz thimbles such that only a single thimble is relevant to the partition function. It can be shown that observables measured in the original and modified theories are connected by a simple identity. We exemplify that our method works well in a toy model.
}
\maketitle
\section{Introduction}\label{intro}
The Monte Carlo method is one of the most important technique to explore nonperturbative physics.
It is based on the fact that the Boltzmann factor $\mathrm{e}^{-S}$ can be regarded as a probability distribution function. Here, $S$ denotes the total action.
In other words, $\mathrm{e}^{-S}$ should be real and positive.
On the other hand, many interesting models in physics have a complex action, and this causes the sign problem.
For instance, the total action of the QCD at finite density becomes complex due to the fermion determinant.
The phase diagram in the finite density region is not understood yet since the standard Monte Carlo simulation suffers from the sign problems. 

In order to overcome the sign problem, the method of Lefschetz thimble has been studied~\cite{Cristoforetti:2012su, Cristoforetti:2013wha, Mukherjee:2014hsa, Fujii:2013sra, Fujii:2015bua, Aarts:2013fpa, Aarts:2014nxa, Kanazawa:2014qma, Tanizaki:2014tua, Tanizaki:2014xba, Tanizaki:2015rda, Alexandru:2015xva, DiRenzo:2015foa, DiRenzo:2017igr}.
It is known that a partition function defined on the real space $\real^n$ can be decomposed into an integral on a set of curved manifolds in the complex space $\complex^n$~\cite{Pham:1983}. These manifolds which are referred to as the Lefschetz thimbles are characterized by the holomorphic gradient flow on which the imaginary part of an action is constant. Basically, the sign problem appearing in the real space becomes mild on the Lefschetz thimbles due to the constancy of the imaginary part of the action. However, even if the thimble decomposition is performed, the sign problem is not completely solved since each integral on a thimble has a global sign factor $\mathrm{e}^{-i\Im S(z_\sigma)}$, where $z_\sigma$ is a fixed point of the flow equation. The interference between thimble integrals could be problematic if a large number of Lefschetz thimbles contribute to the partition function.

In this study, we propose a way to avoid the interference of the global sign factors.
Our strategy consists of two steps. 
In the first step, we modify the partition function such that it is rewritten as an integral on a single Lefschetz thimble.
In the second step, we reconstruct the expectation value defined with the original (unmodified) partition function from that with the modified partition function through a simple identity.
This identity is studied first in our previous paper~\cite{Tsutsui:2015tua} for the cosine model, which is obtained as a special case of the one site U(1) link model.

The outline of this paper is as follows.
In Sec.~\ref{sec-1}, we review the thimble decomposition of the partition function.
We also point out the practical and conceptual difficulty of this technique.
In Sec.~\ref{sec-2}, we give a key identity which plays a central role in this study.
In Sec.~\ref{sec-3}, we demonstrate the modification of thimble structure for a simple toy model.
Section~\ref{summary} is devoted to discussions and summary.

\section{Thimble decomposition}\label{sec-1}
In this section,
we give a brief introduction of the thimble decomposition of partition functions.
We assume that the partition function has a following form:
\begin{align}
Z_f = \int_{D} dx f(x) \mathrm{e}^{-S_{\rm q}(x)},
\label{Zf}
\end{align}
where $f(x)$ is a complex-valued function defined on $x\in\real$ 
and $S_{\rm q}(x)$ is a real-valued action.
$D$ is an integration domain on a real axis.
Throughout this paper, we consider 1-dimensional integral (0-dimensional field theory), for simplicity.
We remark that many interesting models like chiral random matrix models and QCD have partition functions whose form is similar to Eq.~\eqref{Zf}.
More concretely, $f(x)$ in Eq.~\eqref{Zf} mimics the fermion determinant.
 
By exponentiating $f(x)$ in Eq.~\eqref{Zf}, we get
\begin{align}
Z_f = \int_{D} dx \mathrm{e}^{-\lp S_{\rm q}(x) - \log f(x) \rp}.
\end{align}
Clearly seen from this expression, 
the total action $S(x)\equiv S_{\rm q}(x) - \log f(x)$ is complex 
unless $f(x)$ takes real and positive values. 
The complex nature of the total action $S(x)$ leads to the sign problem.

The thimble decomposition is a beautiful framework which improves the oscillatory behavior of the total action.
In general, the partition function can be decomposed as follows~\cite{Pham:1983}:
\begin{align}
Z_f = \int_D dx e^{-S(x)} = \sum_\sigma n_\sigma e^{-i \Im S(z_\sigma)} \int_{\calJ_\sigma} dz e^{-\Re S(z)} .
\label{decomposition}
\end{align}
Here, $z_\sigma$ denotes the complex saddle point defined by $\frac{\del S(z)}{\del z}|_{z=z_\sigma} = 0$,
and $\sigma$ is a label of each saddle point.
The new integration path appearing in the right hand side $\calJ_\sigma$ is the steepest decent path, or so-called Lefschetz thimble, which are defined by the holomorphic flow equation:
\begin{align}
\frac{\del z}{\del t} = + \overline{ \lp \frac{\del S(z)}{\del z} \rp} ,
\label{thimble}
\end{align}
where $\bar{z}$ is a complex conjugate of $z$.
We also introduce the dual of the Lefschetz thimble which is defined by the following flow equation:
\begin{align}
\frac{\del z}{\del t} = - \overline{ \lp \frac{\del S(z)}{\del z} \rp} .
\label{anti-thimble}
\end{align}
This manifold corresponds to the steepest ascent path.
Coefficient $n_\sigma$ in Eq.~\eqref{decomposition} is the intersection number of the original integration path and the steepest ascent path.
If the intersection number is non-zero, we refer to the thimble $\calJ_\sigma$ as a relevant thimble.

On each Lefschetz thimble, 
the oscillatory behavior of the total action is well controlled because its imaginary part $\Im S(z)$ is constant along the thimble.
Thus, the sign factor $e^{-i\Im S(z_\sigma)}$ is factorized out of the integral.

In principle, the Monte Carlo integration of partition function based on the thimble decomposition can be performed by the following procedure:
\renewcommand{\labelenumi}{(\alph{enumi}).}
\begin{enumerate}
	\item Find all saddle points $\{z_\sigma\}$ in the complex plane.
	\item Compute the intersection number $n_\sigma$ for all saddle points.
	\item Perform Monte Carlo sampling with the weight $e^{-\Re S + \log \det J_\sigma}$, where $J_\sigma$ is a Jacobian of the integration measure on thimbles.
\end{enumerate}
Of course, these procedure can be carried out easily for a one dimensional integral~Eq.\eqref{Zf}.
However, it seems hopeless to do that for field theories for the following three reasons.
First, it is quite difficult to find all saddle points (or configurations in a field theory) and compute the corresponding intersection numbers.
Second, usual Monte Carlo simulation does not work if there are several relevant thimbles. Because the thimbles are topologically disconnected, there are infinitely large potential barriers between thimbles.
Third, the sign factor $e^{-i\Im S(z_\sigma)}$ can cause the sign problem if the number of relevant thimbles is large.

The first issue will be solved by considering not only Lefschetz thimbles but also more general manifolds. In~\cite{Alexandru:2015sua, Alexandru:2016gsd, Alexandru:2017czx, Alexandru:2017lqr, Tanizaki:2017yow, Nishimura:2017vav}, a class of manifolds parametrized by the flow time is discussed. Since these manifolds are automatically determined by solving~Eq.\eqref{anti-thimble} with finite flow times, one does not need the locations of complex saddle points explicitly. General manifolds are also studied in~\cite{Mori:2017nwj, Mori:2017pne}, where the manifolds are determined such that a cost function is minimized.
The second issue is closely related to the first issue because one may find an optimal manifold which has finite, but not so large potential barriers by considering general manifolds. One of familiar methods to perform the Monte Carlo sampling on such manifolds is tempering method (replica exchange method)~\cite{Fukuma:2017fjq, Alexandru:2017oyw}.
The objective of the present paper is to propose an approach to solve the third issue.
As we claim in~Sec.~\ref{intro}, we consider a modification of the partition function such that it is written as an integral on a single Lefschetz thimble. Once we obtain such a partition function, the interference of the global sign factors never occurs.

\section{Modification of partition functions}\label{sec-2}
In this section, we derive the key identity which plays a central role in this study.
First, we define the expectation value of an observable $\calO(x)$ as
\begin{align}
\Braket{\calO}_f &\equiv \frac{1}{Z_f} \int_{D} dx \calO(x) f(x) \mathrm{e}^{-S_{\rm q}(x)}.
\label{Of}
\end{align}
In particular, for the quenched model $f(x)\equiv 1$, we simply write
\begin{align}
Z \equiv Z_1 = \int_{D} dx \mathrm{e}^{-S_{\rm q}(x)},
\quad
\Braket{\calO} \equiv \Braket{\calO}_1.
\end{align}
By the definitions, the observable $\calO(x)$ obeys
\begin{align}
\Braket{f}\Braket{\calO}_f = \Braket{f\calO}.
\end{align}
By using the identity, we find the following relation for two arbitrary complex-valued functions $f(x)$ and $g(x)$ and the observable $\calO(x)$:
\begin{align}
{\Braket{f}}\Braket{\calO}_f + {\Braket{g}}\Braket{\calO}_g
&=
{\Braket{f+g}}\Braket{\calO}_{f+g}.
\end{align}
If $\Braket{f}\neq0$, we obtain
\begin{align}
\Braket{\calO}_f
=
\Braket{\calO}_{f+g}
+
\lp
\Braket{\calO}_{f+g} - \Braket{\calO}_{g}
\rp
\frac{\Braket{g}}{\Braket{f}}.
\label{id}
\end{align}
This relation is presented first in~\cite{Tsutsui:2015tua}.
This identity shows that one can obtain $\Braket{\calO}_f$ by computing the observables in other models $Z_g$ and $Z_{f+g}$.
Thus, the central issue is to find an optimal $g(x)$ such that the cancellation of the global phase factors does not occur in the two models $Z_g$ and $Z_{f+g}$.
Therefore, it is necessary that $Z_g$ and $Z_{f+g}$ consist of only one Lefschetz thimble.
We note that $Z_{f+g}$ is obtained by modifying the original partition function.

\section{Gaussian model}\label{sec-3}
Here, we demonstrate how to modify the thimble structure of the original model $Z_f$ by adding a function $g(x)$ with a simple Gaussian integral.
We shall consider the partition function as
\begin{align}
Z_f=\int_{-\infty}^{\infty} dx  f(x;\alpha) e^{-x^2/2}, \quad f(x;\alpha)=(x+i\alpha)^2,
\label{Gaussian_model}
\end{align}
where $\alpha$ is a positive real parameter. 
The total action $S(x)=x^2/2-\log f(x; \alpha)$ is complex when $\alpha \neq 0$,
and it causes the sign problem. 
The analytic expression of an observable in this model can be easily obtained. 
For example, the expectation value of $x^2$ is given by
\begin{align}
\braket{x^2}_f
=
\frac{3-\alpha^2}{1-\alpha^2}. \label{x2}
\end{align}

\begin{figure}[thb]
	\centering
	\includegraphics[width=6cm,clip]{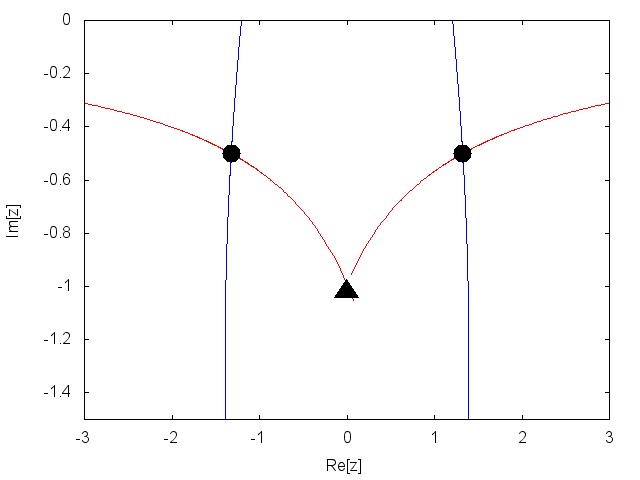}
	\includegraphics[width=6cm,clip]{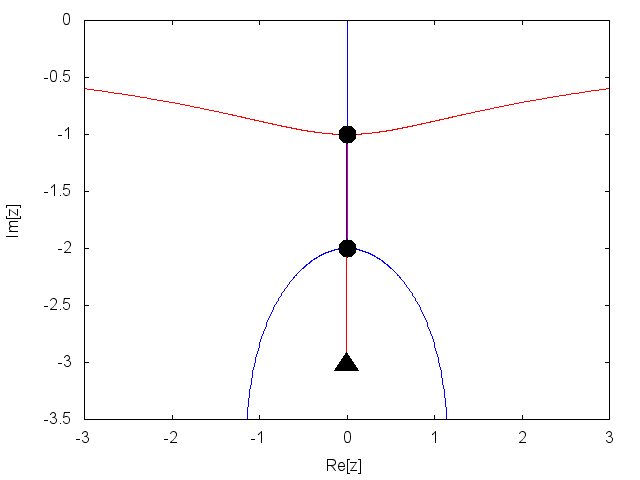}
	\caption{The thimble structures of the Gaussian model for $\alpha=1.0$ (left) and $\alpha=3.0$ (right). Red and blue lines represent the thimbles $\calJ_\sigma$ and its dual $\calK_\sigma$, respectively. The circles and triangles show the location of the fixed points and zero points of $f(z)$.}
	\label{thimble_org}
\end{figure}

In Fig.~\ref{thimble_org}, we show the thimble structure of~Eq.~\eqref{Gaussian_model} for $\alpha = 1.0$, and 3.0, respectively.
The red and blue solid lines represent the Lefschetz thimble and its dual which are one dimensional curves on the complex plane.
The circle and triangular points represent the saddle point and the zero of $f(z)$.
The saddle points are given by
\begin{align}
z_\sigma = -\frac{i\alpha}{2} \pm \frac{\sqrt{8-\alpha^2}}{2}.
\end{align}
While the $f(x)$ is a quadratic polynomial, it has an only one zero point:
\begin{align}
\zeta = -i\alpha.
\end{align}
If $\alpha=2\sqrt{2}$, the saddle point is degenerated.
We find that the Gaussian model has a single-thimble structure for $\alpha > 2\sqrt{2}$.

\begin{figure}[thb]
	\centering
	\includegraphics[width=8cm,clip]{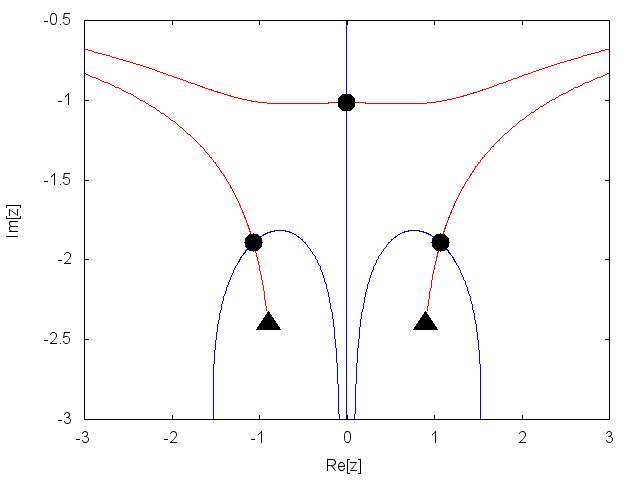}
	\caption{The thimble structure of the modified Gaussian model for $\alpha=1.5, \, \beta=3.3$.}
	\label{thimble_mod}
\end{figure}

In order to use the identity Eq.~\eqref{id}, both $Z_g$ and $Z_{f+g}$ should have the single-thimble structure.
A straight forward way to satisfy these requirements is choosing $g(x)$ as
\begin{align}
g(x;\beta)=f(x;\beta)=(x+i\beta)^2,
\end{align}
with $\beta > 2\sqrt{2}$.
When we employ this function, the modification of the Gaussian model is defined as
\begin{align}
Z_{f+g}=\int_{-\infty}^{\infty} dx \lp f(x; \alpha) + g(x; \beta) \rp e^{-x^2/2}.
\label{mod_Gaussian_model}
\end{align}
The typical thimble structure of $Z_{f+g}$ is shown in Fig.~\ref{thimble_mod}.
Due to the addition of $g(z)$, there are two non-degenerated zero points and three saddle points in general.
Thus, the topology of the thimbles is changed from the original one.
We also find that if $\beta \gtrsim 3.0$, the modified model consists of only one thimble for all $\alpha \geq 0$.

\begin{figure}[thb]
	\centering
	\includegraphics[width=8cm,clip]{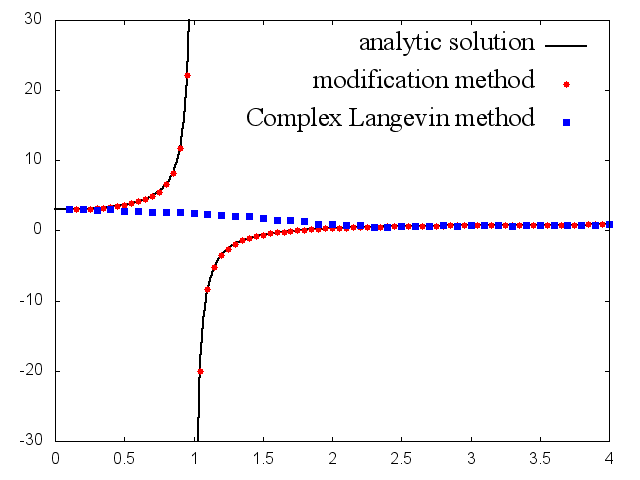}
	\caption{The expectation value $\braket{x^2}$ as a function of $\alpha$. The solid line represents the analytic result~\eqref{x2}.}
	\label{obs}
\end{figure}

We compute the expectation values $\braket{x^2}_g$ and $\braket{x^2}_{f+g}$ by thimble integration for $\beta = 3.3$, and reconstruct the expectation value of the original Gaussian model $\braket{x^2}_f$ via Eq.~\eqref{id}.
Other quantities $\braket{f}$, $\braket{g}$ and $\braket{f+g}$ are computed by the usual Monte Carlo method because the quenched model has no sign problem.
In Fig.~\ref{obs}, we show the numerical results of $\braket{x^2}_f$ as a function of $\alpha$, which is plotted by red circles. 
The solid curve is the analytic result~Eq.~\eqref{x2}.
For comparison, we also show numerical results obtained by the complex Langevin dynamics.
Even where the analytic solution shows quite singular behavior, our numerical result well agrees with that. 

\section{Discussion and Summary}\label{summary}
In this paper, we have proposed a way to modify the thimble structure in order to avoid the effects of global sign factors. 
Here, the modified partition function is defined by adding a complex valued function to the fermion determinant.
By applying this approach to a toy model, we demonstrate how our method works.
We have shown that the identity Eq.~\eqref{id} plays an important role and it enable us to reconstruct the expectation value defined in the original model from that in the modified model.
We have found that our numerical result well agrees with the analytic result even when it has a singularity in the parameter space.

Finally, we comment on the possible objections to our approach.
In this study, we have implicitly assumed that the expectation values $\braket{f}$ and $\braket{g}$ can be computed by Monte Carlo integration. 
Recalling that the former one $\braket{f}$ corresponds to the expectation value of the fermion determinant at large chemical potential, $\braket{f}$ is unlikely calculated in more complicated model.
Other criticism is that our approach seems to be same as the reweighting technique.
Indeed, the ratio $\braket{g}/\braket{f}$ involved in Eq.~\eqref{id} is written as
\begin{align}
\frac{\Braket{g}}{\Braket{f}}
=\frac{\int dx g(x) e^{-S{\rm q}(x)}}{\int dx f(x) e^{-S{\rm q}(x)}}
=\frac{\int dx g(x) e^{-S{\rm q}(x)}}{\int dx \left(\frac{f(x)}{g(x)}\right) g(x)e^{-S{\rm q}(x)}}
=\Braket{\frac{f}{g}}^{-1}_{g}, \label{reweighting_factor}
\end{align}
and this quantity is nothing but the reweighting factor.

We shall remark that in principle, direct computation of $\braket{f}$ is not necessary to our approach.
For instance, if there exists two different functions $g_1$ and $g_2$ which satisfy
\begin{align}
{\Braket{f}}\Braket{\calO}_f + {\Braket{{g_1}}}\Braket{\calO}_{g_1}
&=
{\Braket{f+g_1}}\Braket{\calO}_{f+g_1}, \\
{\Braket{f}}\Braket{\calO}_f + {\Braket{{g_2}}}\Braket{\calO}_{g_2}
&=
{\Braket{f+g_2}}\Braket{\calO}_{f+g_2},
\end{align}
one can solve these equations for $\Braket{\calO}_f$ and $\Braket{f}$.
This idea is slightly generalized and demonstrated for the Gaussian model in our recent paper~\cite{Doi:2017gmk}.
Moreover, this approach works well even when the reweighting factor is small.

We will also explore thimble structure of the massive Thirring model and its modifications.
Since the Thirring model has good analytical properties, studies on this model will give us insights into the theoretical aspects of the modification of partition functions.
The detail of these studies will be reported elsewhere.

\bibliography{Lattice2017_173_SHOICHIROTSUTSUI}

\begin{thebibliography}{27}

\bibitem{Cristoforetti:2012su}
M.~Cristoforetti, F.~Di~Renzo, L.~Scorzato (AuroraScience), Phys. Rev.
  \textbf{D86}, 074506 (2012), \texttt{1205.3996}

\bibitem{Cristoforetti:2013wha}
M.~Cristoforetti, F.~Di~Renzo, A.~Mukherjee, L.~Scorzato, Phys. Rev.
  \textbf{D88}, 051501 (2013), \texttt{1303.7204}

\bibitem{Mukherjee:2014hsa}
A.~Mukherjee, M.~Cristoforetti, Phys. Rev. \textbf{B90}, 035134 (2014),
  \texttt{1403.5680}

\bibitem{Fujii:2013sra}
H.~Fujii, D.~Honda, M.~Kato, Y.~Kikukawa, S.~Komatsu, T.~Sano, JHEP
  \textbf{10}, 147 (2013), \texttt{1309.4371}

\bibitem{Fujii:2015bua}
H.~Fujii, S.~Kamata, Y.~Kikukawa, JHEP \textbf{11}, 078 (2015), [Erratum:
  JHEP02,036(2016)], \texttt{1509.08176}

\bibitem{Aarts:2013fpa}
G.~Aarts, Phys. Rev. \textbf{D88}, 094501 (2013), \texttt{1308.4811}

\bibitem{Aarts:2014nxa}
G.~Aarts, L.~Bongiovanni, E.~Seiler, D.~Sexty, JHEP \textbf{1410}, 159 (2014),
  \texttt{1407.2090}

\bibitem{Kanazawa:2014qma}
T.~Kanazawa, Y.~Tanizaki, JHEP \textbf{1503}, 044 (2015), \texttt{1412.2802}

\bibitem{Tanizaki:2014tua}
Y.~Tanizaki, Phys. Rev. \textbf{D91}, 036002 (2015), \texttt{1412.1891}

\bibitem{Tanizaki:2014xba}
Y.~Tanizaki, T.~Koike, Annals Phys. \textbf{351}, 250 (2014),
  \texttt{1406.2386}

\bibitem{Tanizaki:2015rda}
Y.~Tanizaki, Y.~Hidaka, T.~Hayata, New J. Phys. \textbf{18}, 033002 (2016),
  \texttt{1509.07146}

\bibitem{Alexandru:2015xva}
A.~Alexandru, G.~Basar, P.~Bedaque, Phys. Rev. \textbf{D93}, 014504 (2016),
  \texttt{1510.03258}

\bibitem{DiRenzo:2015foa}
F.~Di~Renzo, G.~Eruzzi, Phys. Rev. \textbf{D92}, 085030 (2015),
  \texttt{1507.03858}

\bibitem{DiRenzo:2017igr}
F.~Di~Renzo, G.~Eruzzi (2017), \texttt{1709.10468}

\bibitem{Pham:1983}
F.~Pham, Proc. Symp. Pure Math. \textbf{40}, 319 (1983)

\bibitem{Tsutsui:2015tua}
S.~Tsutsui, T.M. Doi, Phys. Rev. \textbf{D94}, 074009 (2016),
  \texttt{1508.04231}

\bibitem{Alexandru:2015sua}
A.~Alexandru, G.~Basar, P.F. Bedaque, G.W. Ridgway, N.C. Warrington, JHEP
  \textbf{05}, 053 (2016), \texttt{1512.08764}

\bibitem{Alexandru:2016gsd}
A.~Alexandru, G.~Basar, P.F. Bedaque, S.~Vartak, N.C. Warrington, Phys. Rev.
  Lett. \textbf{117}, 081602 (2016), \texttt{1605.08040}

\bibitem{Alexandru:2017czx}
A.~Alexandru, P.~Bedaque, H.~Lamm, S.~Lawrence (2017), \texttt{1709.01971}

\bibitem{Alexandru:2017lqr}
A.~Alexandru, G.~Basar, P.F. Bedaque, G.W. Ridgway, Phys. Rev. \textbf{D95},
  114501 (2017), \texttt{1704.06404}

\bibitem{Tanizaki:2017yow}
Y.~Tanizaki, H.~Nishimura, J.J.M. Verbaarschot (2017), \texttt{1706.03822}

\bibitem{Nishimura:2017vav}
J.~Nishimura, S.~Shimasaki, JHEP \textbf{06}, 023 (2017), \texttt{1703.09409}

\bibitem{Mori:2017nwj}
Y.~Mori, K.~Kashiwa, A.~Ohnishi (2017), \texttt{1709.03208}

\bibitem{Mori:2017pne}
Y.~Mori, K.~Kashiwa, A.~Ohnishi (2017), \texttt{1705.05605}

\bibitem{Fukuma:2017fjq}
M.~Fukuma, N.~Umeda, PTEP \textbf{2017}, 073B01 (2017), \texttt{1703.00861}

\bibitem{Alexandru:2017oyw}
A.~Alexandru, G.~Basar, P.F. Bedaque, N.C. Warrington, Phys. Rev. \textbf{D96},
  034513 (2017), \texttt{1703.02414}

\bibitem{Doi:2017gmk}
T.M. Doi, S.~Tsutsui (2017), \texttt{1709.05806}

\end{thebibliography}

\end{document}